\documentclass[12pt,preprint]{aastex}
\input psfig.sty
\shortauthors{F. Frontera et~al.}
\shorttitle{GRB011211 with BeppoSAX}

\def\xmm{{em XMM--Newton}}
\def\sax{{\em BeppoSAX\/}}   
\def\xmm{{\em XMM-Newton\/}}
\def\etal{{\it et al. }}

\def\ergcm{\mbox{ erg cm$^{-2}$}}

\begin{document}

\title{The prompt X--ray emission of GRB011211: possible 
evidence of a transient absorption feature}

\author{F. Frontera\altaffilmark{1,2},
L.~Amati\altaffilmark{2},
J.J.M.~in 't Zand\altaffilmark{3},
D. Lazzati\altaffilmark{4},
A. K\"onigl\altaffilmark{5},
M.~Vietri\altaffilmark{6},
E.~Costa\altaffilmark{7},
M.~Feroci\altaffilmark{7},
C.~Guidorzi\altaffilmark{1},
E.~Montanari\altaffilmark{1},
M.~Orlandini\altaffilmark{2},
E.~Pian\altaffilmark{8},
L.~Piro\altaffilmark{7}
}

\altaffiltext{1}{Physics Department, University of Ferrara, Via Paradiso
 12, 44100 Ferrara, Italy; frontera@fe.infn.it}

\altaffiltext{2}{Istituto Astrofisica Spaziale e Fisica Cosmica, section of 
Bologna, CNR, Via Gobetti 101, 40129 Bologna, Italy}

\altaffiltext{3}{Space Research Organization in the Netherlands,
 Sorbonnelaan 2, 3584 CA Utrecht, The Netherlands}

\altaffiltext{4}{Institute of Astronomy, University of Cambridge, Madingley Road, 
Cambridge CB3 0HA, UK}

\altaffiltext{5}{Department of Astronomy and Astrophysics, and Enrico Fermi 
Institute, University of Chicago, 5640 South Ellis Avenue, Chicago, IL60637, 
USA}

\altaffiltext{6}{Scuola Normale Superiore, Pisa, Italy}

\altaffiltext{7}{Istituto Astrofisica Spaziale e Fisica Cosmica, CNR, Via 
Fosso del Cavaliere, 00133 Roma, Italy}

\altaffiltext{8}{Osservatorio Astronomico di Trieste, INAF, Trieste, Italy}


\begin{abstract}
We report on observation results of the prompt X-- and $\gamma$--ray
emission from GRB011211. This event was detected with the 
Gamma-Ray Burst Monitor and one of the Wide Field Cameras aboard the
\sax\ satellite.  The optical counterpart of the GRB was soon
identified and its redshift determined ($z = 2.140$), while with the
\xmm\ satellite, the X--ray afterglow emission was detected. Evidence of 
soft X--ray emission lines was reported by Reeves et al. (2002), but
not confirmed by other authors. In investigating the spectral
evolution of the prompt emission we find the possible evidence of a
transient absorption feature at $6.9^{+0.6}_{-0.5}$ keV during the
rise of the primary event. The significance of the feature is derived
with non parametric tests and numerical simulations, finding a chance
probability which ranges from $3\times 10^{-3}$ down to $4\times
10^{-4}$. The feature shows a Gaussian profile and an equivalent width
of $1.2^{+0.5}_{-0.6}$~keV.  We discuss our results and their possible
interpretation.

\end{abstract}


\section{Introduction}
\label{s:intro}

The nature of the progenitors of celestial Gamma--Ray Bursts (GRBs) is
a still open issue. Collapse of massive fast rotating stars (hypernova
model, e.g., Paczynski 1998\nocite{Paczynski98}) or delayed collapse
of a rotationally stabilized neutron star (supranova model, Vietri \&
Stella 1998\nocite{Vietri98}) are among the favoured scenarios for the
origin of these events.  Both models predict that the pre-burst
environment consists of a high density gas, due to strong winds from
the massive progenitor in the case of a hypernova or to a substantial
enrichment of heavy elements by a previous supernova explosion (SN) in
the case of the supranova model (Lazzati et al. 1999\nocite{Laz99};
B\"ottcher 2000\nocite{B00}; Weth et al. 2000\nocite{Weth00}). The
recent discovery, from the spectral evolution of the optical afterglow
emission from GRB030329, of a direct connection between this burst and
the Type Ic energetic supernova (hypernova) SN2003dh (e.g., Stanek et
al. 2003\nocite{Stanek03}, Hjorth et al.  2003\nocite{Hjorth03}) seems
to point toward the hypernova model, even if the strict simultaneity
of the two events has not been fully proved. It is however not yet
clear whether all GRBs are connected with energetic supernovae,
lacking in many cases evidence of supernova spectroscopic features
and/or a 'supernova bump' in the optical emission of GRB afterglows.
On the other hand, from the current estimates of the opening angles of
GRB jets, the expected GRB rates, when compared with those of SNe Type
Ib/c, suggest that less than 1\% of such SNe might be accompanied by a
GRB \cite{Frail01}. The case of SN2002ap (e.g., Wang et al.
2003\nocite{Wang03}) is an outstanding example of Type Ic hypernova
which has no simultaneous detected GRB event associated with it.

The presence of a post-supernova environment can be tested from the
study of the burst X--ray spectrum, which should show a low-energy
cutoff and/or absorption features due to elements in the circumburst
material left by the previous supernova explosion. Either the
progressive photoionization of the neutral gas by the GRB photons
(e.g., B\"ottcher et al.  1999\nocite{Boettcher99}; Lazzati \& Perna
2002\nocite{LazPer02}) or the electron density decrease or the
electron temperature increase of an already almost photoionized medium
\cite{Lazzati01} would make the features transient.

In fact a transient absorption feature at $3.8 \pm 0.3$~keV has been  
detected in the prompt emission X--ray spectrum of GRB990705
\cite{Amati00}.  The feature is present in the first, rising part
of the burst profile and disappears thereafter.  Interpreted by Amati
et al. (2000) as a cosmologically redshifted K edge due to neutral Fe
around the GRB location, a GRB redshift of $0.86\pm 0.17$ was derived
and later confirmed by optical spectroscopy of the associated host
galaxy ($z_{opt}$ = 0.84, Le Foc'h et al. 2002\nocite{Lefloch02}).
With this assumption, the Iron relative abundance with respect to the
solar one was derived, ${\rm Fe/Fe}_\odot = 75 \pm 19$, which is
typical of a supernova explosion environment.  An alternative
explanation was given by Lazzati et al. (2001)\nocite{Lazzati01}, who
assumed that the feature is an absorption line due to resonant
scattering of GRB photons on H-like Iron (transition 1s-2p, $E_{rest}
= 6.927$ keV). Also in this case the derived redshift is consistent
with that of the host galaxy and the line width is interpreted as due
to the outflow velocity dispersion (up to ~0.1c) of the material,
which should have a Fe relative abundance of $\sim 10$ with respect to
the solar one. In both scenarios, the observed feature points to the
presence of an iron-rich environment left by a recent supernova
explosion.

GRB011211 was localized with the \sax\ WFCs
\cite{Gandolfi01} and promptly followed up with both the \xmm\
satellite and ground based optical/radio telescopes.  In the optical
band, a counterpart was soon discovered \cite{Grav01,Bloom01} and its
redshift determined ($z = 2.140 \pm 0.001$; Fruchter et
al. 2001\nocite{Fruchter01}, Holland et
al. 2002\nocite{Holland02}). In the X--ray band, Santos--Lleo et
al. (2001)\nocite{Santos01} reported the discovery of afterglow
emission with \xmm, while Reeves et al. (2002,
2003)\nocite{Reeves02,Reeves03}, analyzing the \xmm\ data, reported
the evidence of five fading emission lines which were consistent with
blue shifted $K\alpha$ lines from Mg XI, Si XIV, S XVI, Ar XVIII and
Ca XX outflowing at a velocity of $\sim 0.1 c$. However the
significance level of this detection was questioned. Rutledge and Sako
(2003)\nocite{Rutledge03}, examining the same data analyzed by Reeves
et al., found that the claimed lines would not be discovered in a
blind search, showing, with Monte Carlo simulations, that these
features would be observed in 10 percent of featureless spectra with
the same signal-to-noise ratio. Also Borozdin \& Trudolyubov
(2003)\nocite{Borozdin03} questioned the reality of these features.

In the context of a systematic investigation \cite{Frontera04} of the
spectral evolution of all GRBs jointly detected with the Wide Field
Cameras (WFCs, Jager et al. 1997) and Gamma Ray Burst Monitor (GRBM,
Frontera et al. 1997\nocite{Frontera97}, Costa et al.
1998\nocite{Costa98}) aboard the \sax\ satellite, we have analyzed the
prompt emission of this burst. Surprisingly we find a marginal
evidence of an absorption feature which is visible only during the
rise of the event.  If true, this is the second case of a transient
feature after that observed from GRB990705.  In this paper we report
the results of our findings and their possible interpretation.

\section{\sax\ observations} 
\label{s:obs}

GRB011211 was detected with the \sax\ WFC No. 1 and GRBM
\cite{Frontera01} on 2001 December 11 at 19:09:21 UT. Since a
follow-up with \xmm\ was promptly scheduled for this GRB, no
observation with the \sax\ Narrow Field Instruments was performed.
The GRB position was determined with an error radius of $2'$ (99\%
confidence level) and was centered at $\alpha_{2000}\,=\, 11^{\rm
h}15^{\rm m}16^{\rm s}$, and $\delta_{2000}\, =\, -21^\circ55'44''$
\cite{Gandolfi01}.

Data available from the GRBM include two 1~s ratemeters in two energy
channels (40--700~keV and $>$100~keV), 128~s count spectra
(40--700~keV, 225 channels) and high time resolution data (down to
0.5~ms) in the 40--700~keV energy band. The WFCs (energy resolution
$\approx$ 20\% at 6~keV) were operated in normal mode with 31 channels
in 2--28~keV and 0.5~ms time resolution.  The burst direction was
offset by 9.1$^\circ$ with respect to the WFC axis. With this offset,
the effective area exposed to the GRB was $\approx$~500~cm$^2$ in the
40-700~keV band and 65~cm$^2$ in the 2--28~keV energy band.  The
background in the GRBM energy band was estimated by linear
interpolation using the 250~s count rate data before and after the
burst.  The WFC spectra were extracted through the Iterative Removal
Of Sources procedure (IROS
\footnote{WFC software version 105.108}, e.g.  Jager et al. 1997
\nocite{Jager97}) which implicitly subtracts the contribution of the
background and of other point sources in the field of view.

\begin{figure}[!t]
\epsscale{0.8}
\plotone{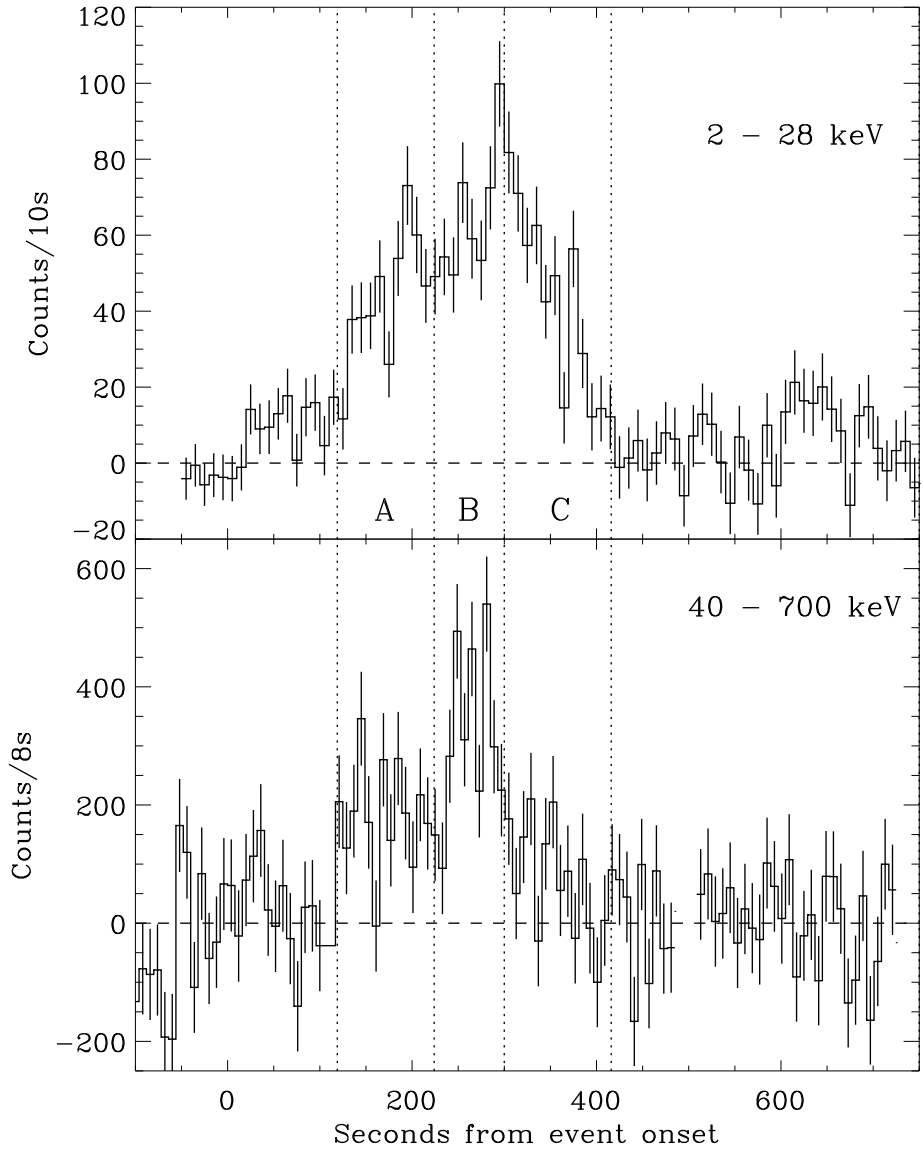}
\vspace{1.5cm}
\caption{Light curve of GRB011211 in two energy bands, 2--28 keV (WFC),
40--700 keV (GRBM), after background subtraction. The zero abscissa corresponds 
to 2001 Dec. 11, 19:04:31 UT. The time slices on which the spectral analysis was 
performed are indicated by vertical dashed lines.}
\label{f:lc}
\end{figure}

Among the gamma--ray bursts simultaneously detected by the \sax\ WFC
and GRBM, GRB 011211 is the longest in both X-- (2--28 keV) and
$\gamma$--rays (40--700~keV).  Figure~\ref{f:lc} shows its light curve
in these energy bands. In X--rays (top panel of Fig.~\ref{f:lc}) it
shows a long rise ($\sim 300$~s) and a shorter decay ($\sim 120$~s).
Secondary peaks are also visible: three during the GRB rise and one
during the burst decay.  In gamma--rays (bottom panel of
Fig.~\ref{f:lc}), the count statistics do not allow a precise
determination of the onset time, which seems to occur later ($\sim
100$ s) than the onset of the X--ray emission. However the gamma--ray
peak is achieved about 20 s earlier than in X--rays and the GRB end
appears to be simultaneous in X-- and gamma--rays. The GRB entire
duration in gamma--rays is $\sim 270$~s.

\section{Spectral analysis and results}
\label{s:results}

\begin{figure}[!t]
\epsscale{0.8}
\plotone{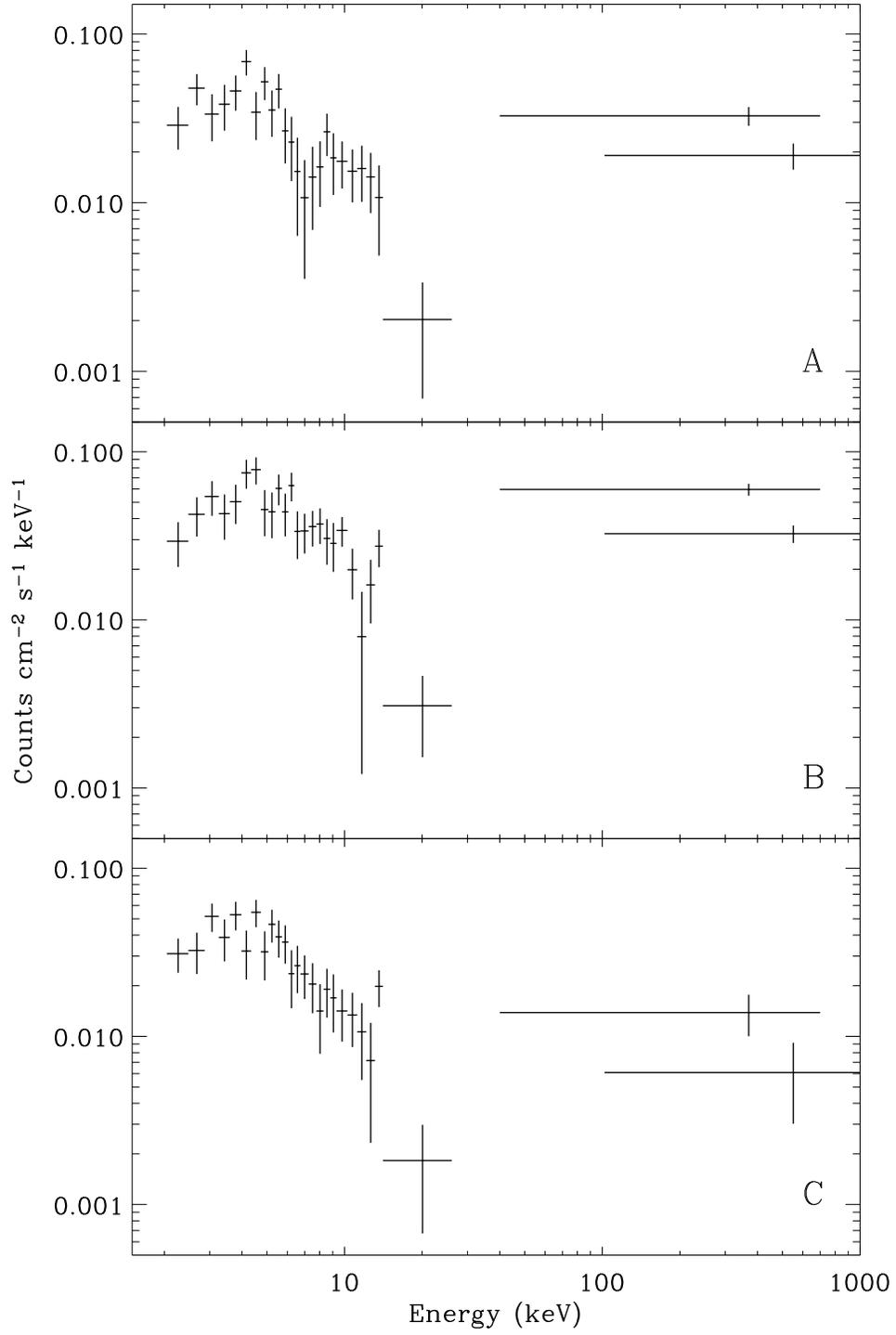}
\vspace{1.5cm}
\caption{Count spectra  of the burst in the time intervals A, B, and C.}
\label{f:sp}
\end{figure}

Following the investigation performed on a sample of \sax\ GRBs
\cite{Frontera00}, we are performing the systematic analysis of all
GRB detected with the \sax\ WFCs and GRBM, in order to study their
spectral evolution in the 2--700 keV energy band.  Our method is that
of subdividing the pulse profile in a certain number of slices of
duration such to derive statistically significant count spectra for
each of them. For GRB011211, we subdivided the light curve in 3
intervals A, B, C (see Fig.~\ref{f:lc}), 100~s, 76~s, and 116~s
duration, respectively.  The derived 2--700 keV count rate spectra are
shown in Fig.~\ref{f:sp}.  The surprise was the shape of the spectrum
in the time interval A, which showed at a glance, between 6 and 8 keV,
a structured depression, which was not apparent in the other two
spectra, which exhibited the typical shape found in many other GRBs.
Subdividing the 100~s interval in two sub-intervals, the count
statistics were very low, but, in spite of that, the depression
marginally appeared in both spectra.  Subdividing the interval B in
two sub-intervals, no trace of the feature was found in either
spectra.
 
\begin{figure}[!t]
\epsscale{0.8}
\plotone{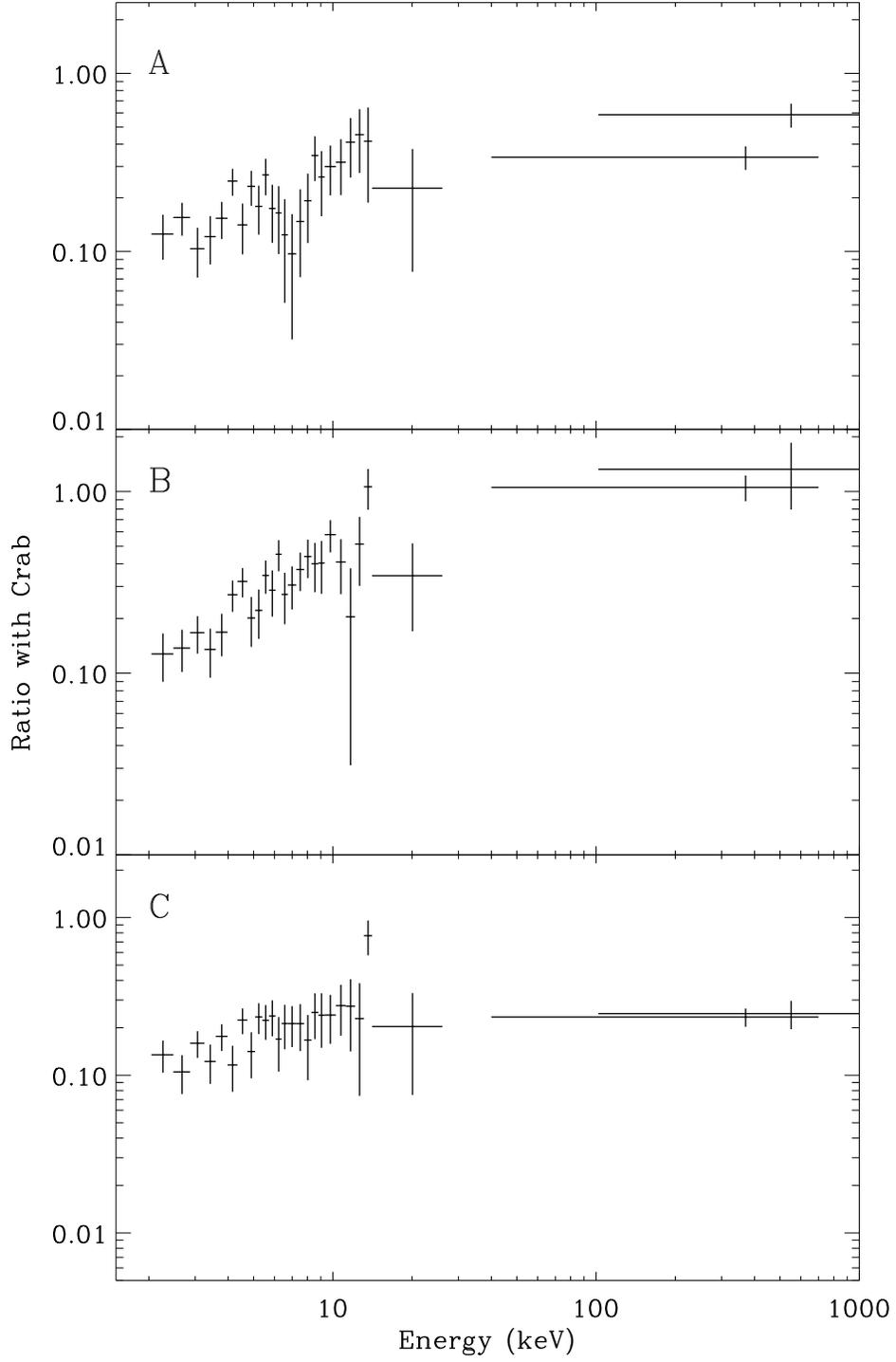}
\vspace{1.5cm}
\caption{Ratio  of the WFC spectra in the time interval A, B and C with the
Crab Nebula spectrum measured by the WFC with the source in same offset as that 
of GRB011211.}
\label{f:crab_ratio}
\end{figure}

The feature in the interval A is also evident (see
Fig.~\ref{f:crab_ratio}) when we perform the ratio of the GRB count
spectra with that of the Crab Nebula, which by chance was observed at
an angular offset similar to that (9.1$^\circ$) of GRB011211.  The
Crab ratio criterion, commonly used to search for cyclotron resonance
features in X-ray pulsars (e.g., Dal Fiume et
al. 2000\nocite{Dalfiume00}), has the advantage of being independent
of the detector response function adopted and thus minimizing the
calibration uncertainties.

We investigated a possible instrumental origin of this depression with
negative results.  The high-voltage, which is monitored every second,
does not show any glitch in the interval A. Given the accuracy of the
readout (which dominates over statistical noise) this excludes gain
changes higher than 0.01\%.  All the ratemeters show the GRB event,
even the ones that measure the illegal events, so there is nothing
that can be concluded from them.  There are no dips or spikes of any
sort, with a typical 3$\sigma$ upper limit of 10\% per second of
measurement (for a count rate of 700 counts s$^{-1}$).  All monitored
quantites at the time of the depression do not show anything out of
the ordinary.  Finally, note that the same procedure was followed to
extract both the spectrum A and the spectra B and C, in which the
feature is not observed.

As for the other events, the joint WFC and GRBM count rate spectra A,
B and C were analyzed by means of the {\sc xspec} software package
\cite{Arnaud96}.  In the fits, allowing the normalization factor of
the GRBM data with respect to the WFC data free to vary, we tested
various continuum models: power-law ({\sc pl}), broken power-law ({\sc
bknpl}) and smoothly broken power-law (or Band law, {\sc bl}, Band et
al. 1993\nocite{Band93}).  The {\sc pl} model provided a good fit only
with a GRBM/WFC normalization factor of 0.3 which is unusually low.
Indeed, from extended flight calibrations with celestial sources
(e.g., Crab Nebula) and using GRBs observed with both the WFC-GRBM and
the CGRO/BATSE experiment (e.g., Fishman et
al. 1994\nocite{Fishman94}), this factor was found to range from 0.8
to 1.3. Thus we excluded the {\sc pl} model.  The {\sc bl} gave a good
fit, but the best fit parameters derived were not constrained by the
data. The best data description and the best parameter constraints
were obtained with a {\sc bknpl}, with best fit parameters reported in
Table~\ref{t:results}. In the fits the GRBM/WFC normalization factor
was fixed to 0.8, but we verified that the fit quality and the {\sc
bknpl} parameter values did not significantly change by using other
values of this factor in the 0.8--1.3 range.  We also took into
account, in the fits, photoelectric absorption of the radiation from
the GRB, finding that the equivalent hydrogen column density $N_{\rm
H}$ was consistent with zero. Thus we assumed a Galactic absorption
along the GRB direction ($N_{\rm H} = 4.2 \times 10^{20}$~cm$^{-2}$,
Dickey \& Lockman 1990\nocite{Dickey90}).

\begin{figure}[!t]
\psfig{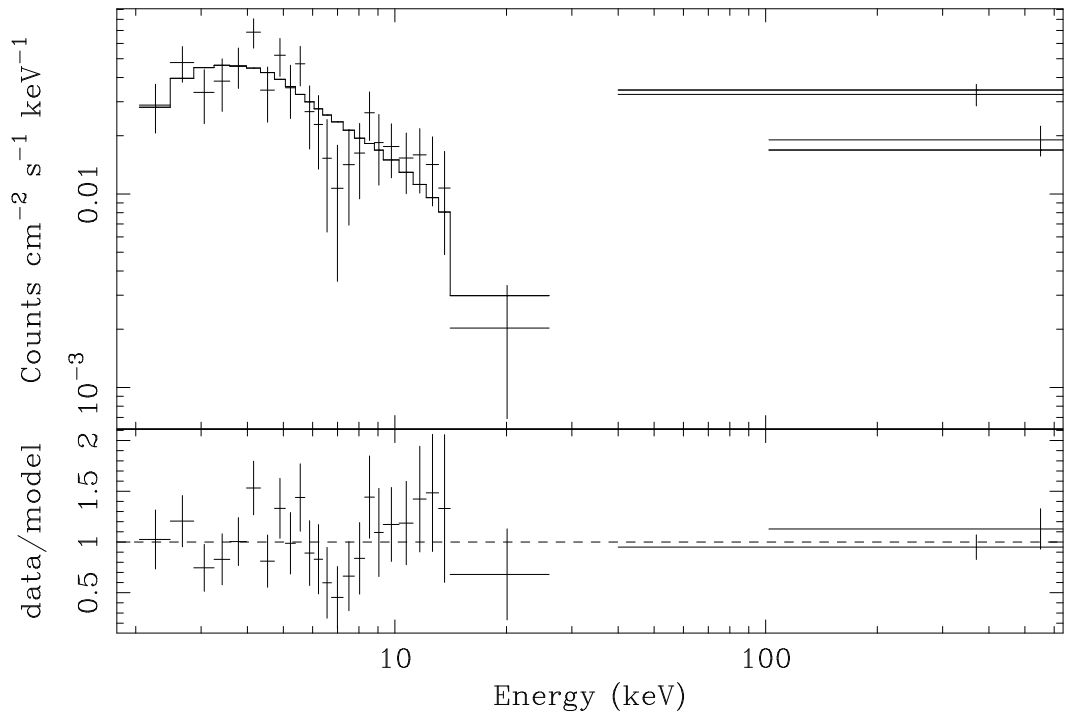}
\vspace{0.5cm}
\caption{{\it Top}: Count spectrum of GRB011211 in the interval A. The step-like
curve represents the best fit with a {\sc bknpl}. {\it Bottom}: Distribution of the
residuals to the model.}
\label{f:residuals}
\end{figure}

Even if, for the interval A, the {\sc bknpl} model already provides an
acceptable fit ($\chi^2/{\rm dof} = 20.4/22$ corresponding to a chance
probability of 0.57, see Table~\ref{t:results}), the distribution of
the residuals from the model (see Fig.~\ref{f:residuals}) still shows
evidence of the structured depression.  The feature is still apparent
in the residuals from the best fits with the other models ({\sc pl}
and {\sc bl}) and is also evident when only the WFC data are fit with
a {\sc pl} model. The used models are those generally used to describe
the prompt continuum emission from GRBs. However we cannot exclude
that a properly tuned two--component continuum model could recover the
feature, even if, on the basis of the results obtained thus far in the
X--ray band ($<$ 10 keV) with \sax\ and with {\it HETE-2} (e.g.,
Barraud et al. 2003\nocite{Barraud03}), this possibility appears much
less probable.

Thus we decided to evaluate from a statistical point of view the
significance of the feature using the residuals from the {\sc bknpl}
best fit model.  Given that the $\chi^2$-test is a global test, which
does not take into account the relative position of the residuals, we
performed a null-hypothesis test to check the independence of the
residuals around the depression (see Fig.~\ref{f:residuals}). For that
we used the run-test (e.g., Bendat and Piersol 1971\nocite{Bendat71}),
a non-parametric test, which is valid independently of the statistics
assumed for the data.  Taking the residuals to the {\sc bknpl} as data
points, we investigated the probability that groups of $n$ consecutive
data points are mutually independent.  Conservatively the $n$ value
was chosen to be the number of energy bins corrersponding to 4 times
the energy resolution of the instrument in correspondence of the
depression region. Thus we considered values of $n$ ranging from 11 to
14 bins.  For a chosen value of $n$, we scanned the entire count
spectrum, by moving the first bin of the group. We found that the
probability that $n$ consecutive residuals from the {\sc bknpl} model
are mutually independent, ranges from about 0.16 to 0.47 when the
energy bins are far from the depression, and from $2.7\times 10^{-3}$
down to $7.2\times 10^{-4}$ (equivalent, in a Gaussian approximation,
to 2.8--3.1 $\sigma$), depending on $n$, when the energy bins cover
the depression region. The probability values obtained in the
depression region for the different values of $n$ are reported in
Table~\ref{t:rtest}.

\begin{figure}[!t]
\psfig{figure=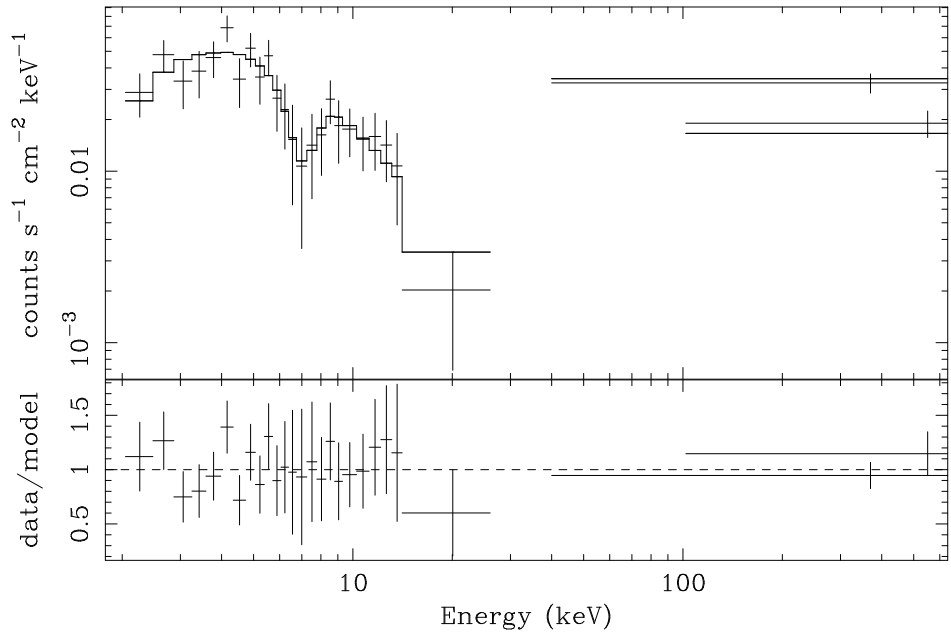,width=14cm,angle=-90}
\vspace{0.5cm}
\caption{{\it Top}: Photon spectrum of GRB011211 in the time interval A. The step-like
curve represents the best fit with the model given by eq.~\ref{e:model}. {\it Bottom}:
Distribution of the residuals to the model.}
\label{f:best_fit_A}
\end{figure}

Stimulated and encouraged by the run-test results, we replaced the
photoelectrically absorbed {\sc bknpl} model with the same law
modified by a negative narrow Gaussian profile, according to the
following equation:
\begin{equation}
N(E) = \left\{ \begin{array}{ll}
       A \exp{(-\sigma N_{\rm H})} E^{-\Gamma_1}\left[1- EW\, G(E; E_L, \sigma_l)\right]  
                               & E\le E_b \\
      A E_b^{\Gamma_2 - \Gamma_1} E^{-\Gamma_2}
            \left[1- EW\, G(E; E_l, \sigma_l)\right]  & E\ge E_b
 \end{array}
\right.
\label{e:model} 
\end{equation}
where $\sigma\,=\,\sigma (E)$ is the photo-electric cross-section of a
gas with cosmic abundance (Morrison \& McCammon 1983), and $E_b$ is
the break energy, A is the normalization factor, $EW$ is the
equivalent width of the line, and
\begin{equation}
G(E; E_l, \sigma_l) = \frac{1}{\sqrt{2\pi \sigma_l}} \exp{\left[\frac{(E - E_l)^2}
{2 \sigma_l^2}\right]}
\end{equation}
is the assumed Gaussian profile, with centroid energy $E_l$ and
standard deviation $\sigma_l$.  A similar model is, e.g., used to
describe cyclotron absorption features from X-ray pulsars (e.g., Soong
et al. 1990\nocite{Soong90}).  The fit results with this new model,
instead of the {\sc bknpl}, are reported in Table~\ref{t:results}. In
the fits $\sigma_l$ is set to 0.1 keV.  Leaving it free to vary, the
best fit is still obtained for $\sigma_l \approx 0.1$~keV with an
upper value of 1.2 keV at 90\% confidence level.  In
Fig.~\ref{f:best_fit_A} we show the A spectrum with the best fit curve
and the residuals of the data from the model. As can be seen, the
depression, has now disappeared ($\chi^2/{\rm dof} = 12.1/20$).

Being aware of the warnings given by Protassov et
al. (2002)\nocite{Protassov02} about the use of the F-test for the
inclusion of an additional term to a fitting function
\cite{Bevington69} when this function is an absorption or emission
feature, in order to evaluate the probability that the parent fitting
function of the spectrum A is actually given by the Eq.~\ref{e:model},
we followed the following approach. Assuming as featureless spectrum
the {\sc bknpl} model that best fits the WFC plus GRBM data (see
Table~\ref{t:results}), by means of {\sc xspec} we generated 10\,000
WFC fake spectra with the same number of energy bins of the measured A
spectrum. These spectra take into account the time duration of the
spectrum A, the response function of the instrument and the Poisson
noise associated to the measured spectrum.  We fit each of these
spectra with a {\sc bknpl} model, obtaining a set of 10\,000
distributions of residuals from the model assumed.  For each of these
residual distributions, we performed a blind run-test, determining the
spectra which have at least one group of $n$ consecutive residuals
from the {\sc bknpl} whose probability of being mutually independent
is less than the probabilities reported in Table~\ref{t:rtest}.  The
fraction $f_r$ of the spectra that satisfy this condition are reported
in Table~\ref{t:rftest}, for values of $n$ in the range from 11 to 14,
as above.

Given that the run-test takes into account the relative position of
the residuals but not their amplitude, we also fit these spectra with
the Eq.~\ref{e:model} model, and, mindful of the Protassov et
al. (2002) warnings, on these spectra we performed not the F-test for
testing the addition of a further component to a fitting function, but
for testing the discrepancy between an assumed fit function and the
parent function.  This test (see Bevington 1969,
p. 195\nocite{Bevington69}) makes use of the probability distribution
of the ratio $F_{12} = \chi^2_{\nu_1}/\chi^2_{\nu_2}$ (or $F_{21} =
1/F_{12}$) where $\chi^2_{\nu_1} = \chi^2/\nu_1$ is obtained from the
fit of the function 1 (in our case the {\sc bknpl} model) to the data,
and $\chi^2_{\nu_2} = \chi^2/\nu_2$ is obtained from the fit of the
function 2 (in our case, Eq.~\ref{e:model}).

In the spectral fits with the Eq.~\ref{e:model}, the $\sigma_l$ was
again frozen to 0.1 keV, and, in order to perform a blind search of a
negative Gaussian feature, for each spectrum we performed 6 fits, each
with a different starting value of the feature centroid. In this way
we determined the fraction $f_t$ of spectra which satisfy the
condition $F_{12}\ge F_{12}^{m}$, where $F_{12}^{m}$ is the value
measured of $F_{12}$, derivable from the $\chi^2$ values reported in
Table~\ref{t:results}.  We also checked the condition $F_{21} \le
F_{21}^{m}$, finding the same result.  The values of $f_t$ versus $n$
are reported in Table~\ref{t:rftest}. As can be seen, the fraction of
simulated spectra which satisfy both tests ranges from $3 \times
10^{-3}$ down to $4 \times 10^{-4}$. These fractions give our best
estimate of the statistical significance of the negative Gaussian
feature observed.

In the above estimates we did not take into account the number of
trials whereby such a feature could be found as a result of a random
fluctuation. Thus, assuming that the range of values found for $f_t$
as chance probability range for a random fluctuation, using the
binomial distribution, we evaluated the probability $P(2)$ that,
observing $n$ GRBs, two of them (GRB990705 and GRB011211) show during
their leading edge an absorption feature.  To determine $n$, from the
entire sample of GRBs detected with both WFC and GRBM we selected
those which show during their leading edge 2--10 keV count spectra
with statistical quality at least similar to that of the spectra in
the time intervals in which the absorption features are apparent
(interval A for GRB011211, interval B for GRB990705).  We found $n =
18$ and thus $2.4 \times 10^{-5} \le P(2) \le 1.3 \times 10^{-3}$,
depending on the chosen value of the chance probability $p$, $4 \times
10^{-4}$ or $3 \times 10^{-3}$, respectively. The low values of $P(2)$
gives us an additional confidence that the feature observed is not a
random fluctuation, at least on the basis of the GRB sample size
obtained with \sax.

Assuming the best fit centroid energy of the negative Gaussian and
$\sigma_l = 0.1$~keV, we also evaluated the $2\sigma$ upper limit to
the feature $EW$ in the time intervals B and C.  These upper limits
and the corresponding $\chi^2$ are also reported in
Table~\ref{t:results}.

We have also tested a K edge model ({\sc edge} in {\sc xspec}) in the
place of the negative Gaussian. In this case the fit improvement is
lower, likely due to the symmetrical shape of the line which does not
favor this model.
 
On the basis of the best fit results, we have derived the energetics
associated with the event. The fluence of the GRB is $S_{\rm 2-10} =
1.1 \times 10^{-6}$~erg cm$^{-2}$~\ergcm\ in the 2--10 keV band, and
$S_{\rm 40-700} = 5.1 \times 10^{-6}$~erg cm$^{-2}$~\ergcm\ in the
40--700 keV band, while the total 2--700 keV fluence is $S_{\rm 2-700}
= 7.5 \times 10^{-6}$~erg cm$^{-2}$.  For comparison, in the time
interval A (100 s duration), the 2--700 keV fluence is $\sim 2.6\times
10^{-6}$~erg cm$^{-2}$.  The peak flux averaged on a 8 s bin is
$F_{\rm 2-10}^p = 1.56 \times 10^{-8}$~erg~cm$^{-2}$~s$^{-1}$ in the
2--10 keV band, and $F_{\rm 40-700}^p = 5.0 \times
10^{-8}$~erg~cm$^{-2}$~s$^{-1}$ in the 40--700 keV interval.  >From
the optical redshift, assuming a standard cosmology with $H_{\rm 0} =
65$~km s$^{-1}$ Mpc$^{-1}$, $\Omega_\lambda = 0.7$ and $\Omega_m =
0.3$, and isotropic emission, the total fluence corresponds to a
released energy of $\sim 3.55 \times 10^{52}$~erg, with 34\% of it
released during the first GRB interval. The released energy in the
time interval A above 7 keV (feature centroid energy) is $1.2 \times
10^{52}$~erg, which corresponds to 98\% of the energy released by the
burst in this interval.

\section{Discussion} 
\label{s:disc}

There is a long history of X-ray lines from GRBs. In the eighties,
evidence of hard X--ray ($>$10 keV) absorption features was reported
by Mazets et al. (1981)\nocite{Mazets81}, Hueter
(1984)\nocite{Hueter84}, and Murakami et al. (1988). The most
significant result was obtained by Murakami et al. (1988) with the
Gamma Burst Detector aboard the {\it Ginga} satellite, with two
transient absorption lines observed during the latest part (5~s
duration) of the GRB880205 leading edge. The line energies were $\sim
19$ and $\sim 38$ keV and their chance probability was evaluated to be
$6 \times 10^{-5}$ \cite{Fenimore88}. Another significant hard X--ray
line detection, which was visible, like the previous one, during the
leading edge of the burst, was later detected with {\it Ginga} from
GRB890929 and reported by Yoshida et al.  (1991)\nocite{Yoshida91}.

All these features at those times had a natural explanation in terms
of cyclotron features from Galactic neutron stars (GNS). Once this
association with GNS resulted to be untenable owing to the isotropic
distribution of GRBs discovered with BATSE, the general belief was
that all these feautures likely were not real. Possible justifications
included inaccurate knowledge of the true response functions of the
instruments used, partly due to the indetermination of the GRB
direction. Another tacit justification was the difficulty of making
these lines at cosmological distances.  No detection of hard X--ray
lines was reported by the BATSE team.

With the advent of the afterglow era, evidence of X--ray lines has
been found both in the GRB afterglow and in the prompt emission.
Concerning the afterglow, emission lines with energy centroids below
10 keV, have been reported from several GRBs (see, e.g., review by
Frontera 2003\nocite{Frontera03a}), most of which pointing to Iron
fluorescence or recombination lines from circumburst
regions. Concerning the GRB prompt emission, as discussed in Section
1, a transient absorption feature was detected during the rise time of
GRB990705 \cite{Amati00}, In that case, the identification of the
absorption feature, either with a K--edge of neutral Iron
\cite{Amati00} or with a resonant scattering line of nearly ionized
Iron \cite{Lazzati01}, led to a redshift coinciding with the host
galaxy redshift independently measured in the optical band
\cite{Lefloch02}. This fact strenghtened the result.

Now we have the possible evidence of another similar feature from
GRB011211. Also in this case, the feature is visible only during the
rise of the primary event.  However, in the case of GRB011211 the
interpretation is not so straightforward.  First of all, the feature,
due to its symmetrical shape, is more consistent with an absorption
line than a K edge. Also the assumption of a cyclotron resonance
feature, which was the assumption for the origin of the {\it Ginga}
lines, appears untenable. If an oriented magnetic field were present
in the emitting region, the magnetic field strength can be derived
from the condition that the magnetic energy density integrated over
the emission region of the prompt emission times a radiative
efficiency factor (likely not smaller than $\sim 0.1$) is at most
equal to the estimated energy released (see
Section~\ref{s:results}). The result is that $B = 10^5$--$10^6$~Gauss.
This appears inconsistent with the magnetic field strength ($\sim
10^{12}$ Gauss) obtained assuming that the line is a cyclotron
resonance feature ($E_{rest} = 21.7^{+1.9}_{-1.6}$~keV after
correcting it for the optically-determined redshift). Neither this
large magnetic field seems to be compatible with the value derived
from the analysis of the afterglow data
\cite{Jakobsson03} using the standard homogeneous fireball model \cite{Sari98}.

The rest-frame energy of the absorption feature is also inconsistent
with any plausible resonant scattering line unless we invoke a blue
shift of the line. Indeed the most abundant elements of typical
interstellar environments or those typical of a supernova explosion
have resonant scattering energies far below the rest--frame energy of
the discovered feature (see, e.g., Kato 1976\nocite{Kato76}). This
implies an absorbing medium with very high outflowing velocities
$v_{out}$.  A lower limit to this velocity is obtained assuming a
resonant scattering of the GRB photons off H--like Ni XXVIII
(rest-frame energy of 8.092 keV). To derive the outflow speed, we
consider that photons seen in the ouflow comoving frame at $\nu=\nu_0$
will be absorbed, where $\nu_0$ is the resonance frequency of the
considered line. The following condition needs therefore to be
satisfied:
\begin{equation}
(1+z)\,\delta\,\nu_{\rm{abs}} = \nu_0
\end{equation}
where $\nu_{\rm{abs}}$ is the observed absorption frequency and
$\delta\equiv[\gamma(1-\beta\cos\theta)]^{-1}$ is the Doppler
factor. In our case, since the photon and the fluid direction
coincide, we have $\cos\theta=-1$ and the equation can be rewritten,
after simple algebra,
\begin{equation}
\frac{1+\beta}{\sqrt{1-\beta^2}}=(1+z)\frac{\nu_{\rm{abs}}}{\nu_0}
\label{eq:ref}
\end{equation}
where $h\nu_0$ is the rest frame energy of the candidate feature.  In
this case we get $\beta=v_{out}/c=0.75$, with a corresponding Lorentz
factor $\gamma=1.5$. Note that in this trans-relativistic regime nor
the $\beta\sim1$ neither the $\gamma \sim1$ approximations can be
adopted. Eq.~\ref{eq:ref} simplifies though to the usual formula
$(1+z)\,\nu_{\rm{abs}}=2\,\gamma\,\nu_0$ in the ultra-relativistic
regime.

\begin{figure}[!t]
\psfig{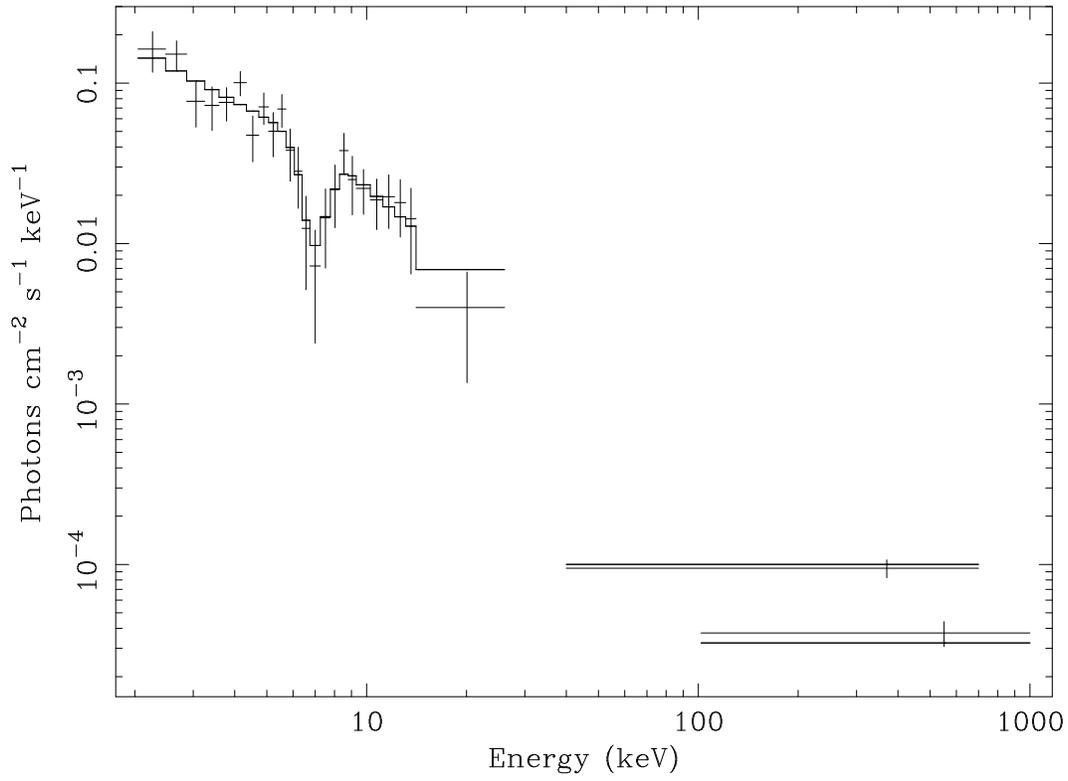}
\vspace{0.5cm}
\caption{{\it Top}: photon spectrum of GRB011211 in the time interval A.  The step-like
curve represents the best fit with the model used in Lazzati et al. (2001) (see 
text).}
\label{f:lazzati}
\end{figure}

Assuming, as proposed by Lazzati et al. (2001) for GRB990705, that the
absorption feature is due to resonant scattering of GRB photons off a
mixture of H--like Fe and Co ($E_{rest} = 6.927$ and $7.518$~keV,
respectively; oscillator strength of 0.416, Kato 1976), the best fit
($\chi^2/{\rm dof} =13.1/20$) is obtained assuming a
super-thermal\footnote{A thermal distribution of outflow velocity
would give a too narrow resonant feature, since $kT_{\rm{Fe}}<10$~keV
in order to have a sizable fraction of non fully ionized iron ions.}
Gaussian distribution of the product $\beta \gamma$ where $\gamma$ is
the bulk Lorentz factor of the expanding medium, with
$\sigma_{\beta\gamma}/(\beta \gamma) = 7.5$\% and negligible Fe K-edge
of the absorbing medium.  The best fit parameters of the model (mean
value of $\beta \gamma$, and column density of the Fe absorbing
material) are reported in Table~\ref{t:results}.  From the best fit
value of $\beta \gamma$, we derive, by using again
Eq.~\ref{eq:ref}, a value of $\beta = 0.80$ and $\gamma=1.65$. The
deconvolved spectrum of the interval A with superposed the best fit
model is shown in Fig.~\ref{f:lazzati}.  The low depth of the Fe K
edge could be the result of the high ionization level and/or the edge
broadening due to Compton scattering
\cite{Ross99}.  The disappearance of the feature after $\sim100$~s is
likely the result of the inhibition of the electron recombination
caused by the increase of the electron temperature as a consequence of
Inverse Compton heating \cite{Lazzati01}.

The high outflowing velocity is a problematic issue. 
The acceleration cannot be due to the interaction with the burst
photons themselves, since the feature centroid does not evolve when
the interval A is analyzed with higher temporal
resolution. A possible explanation is that the absorption
feature is due to outflowing material in the line of sight toward the
jet cone (see also  K\"onigl 2004\nocite{Konigl04}). 

The fit with the physical model allows us to estimate the
distance of the absorber and the amount of absorbing material. From Eq.~12
of Lazzati et al. (2001) we derive:
\begin{equation}
R_{\rm{Abs}} \simeq 5\times10^{16} \quad {\rm cm}
\end{equation}
while the fit yields $N_{\rm{Fe+Co}}\simeq10^{20}$~cm$^{-2}$. Assuming
isotropy, the mass of the absorber results in
$M_{\rm{Fe+Co}}=0.15M_\odot$, the total mass being
$M\sim (60/A_{\rm{Fe}}) M_\odot$ where $A_{\rm{Fe}}$ is the iron (cobalt)
richness in solar units. This estimate implies a moderate iron and
cobalt enriched material in order to keep the total mass to a
reasonable value. Alternatively, one can envisage a clumped material,
with a covering factor much lower than unity, with a dense clump lying
randomly along the line of sight. In addition, it is possible to
derive the total energy involved in the outflow, from the best-fit
average speed. We obtain
$E_{\rm{kin}}=(\gamma-1)Mc^2\sim 5\times10^{55}/[A_{\rm{Fe}}\eta(\Omega/4\pi)]$~erg
where $\eta$ is the covering factor of the clumps and $\Omega$ is the
solid angle within which the material is accelerated to the high speed
obtained from the fit. Reasonable values are $A_{\rm{Fe}}\sim 10$,
$\eta\sim 0.1$ and $\Omega/4\pi\sim 0.1$, resulting in a large amount of
kinetic energy $E_{\rm{kin}}\sim5\times10^{52}$~erg, a value larger
than the typical GRB energy of $\sim10^{51}$~erg \cite{Bloom03}.

Given that any absorption feature in the GRB prompt emission,
specially during its rise, implies that the absorbing material is
external to the fireball, there is only one model, among the proposed
progenitors, which can accommodate this observation. In the Supranova
scenario \cite{Vietri98} a supernova explodes prior to the GRB,
ejecting high speed material around a newly formed rotating
supra-massive neutron star. The GRB is produced when the spin of the
neutron star is reduced and the compact object collapses to a black
hole.  In the time span between the supernova and GRB explosions, the
spindown of the neutron star releases $E_{\rm SD} \sim 10^{53}$ erg,
of which a significant fraction could be in the form of a pulsar-type
wind. As discussed in K\"onigl \& Granot (2002), this energy might be
deposited in a pulsar-wind bubble and used to accelerate the supernova
shell. They in fact argued that the swept-up remnant is likely to
expand anisotropically, with the polar regions of the shell (which can
be identified with the gas responsible for the high-energy absorption
detected in GRB 011211) attaining speeds $\sim c$.\footnote{In the
above picture, the ratio of the distances of the emission and
absorption regions of the remnant, $R_{\rm Em}/R_{\rm Abs}$, would be
approximately equal to the ratio of their respective expansion speeds
($\sim 0.1/0.8$ in this case). Using Eq. (3), this yields $R_{\rm Em}
\sim 6 \times 10^{15}$ cm.}  On the other hand, explaining the
presence of a large amount of high velocity absorbing material in the
immediate vicinity of the burster is extremely problematic for any
model invoking a simultaneous explosion of the burst with the
progenitor star (e.g. Woosley 1993)\nocite{W93}, since the stellar
wind creates an iron poor (solar at most) low-density environment at
the required distance.

Only more sensitive X--ray observations than those performed with
the \sax\ WFCs can enlarge the GRB sample, improve the statistical quality
of the time resolved spectra, and thus test the results 
obtained with \sax\ on the X--ray transient absorption lines.
These observations are expected to be only partly covered by the upcoming satellite 
{\em Swift} (e.g., Krimm 2004\nocite{Krimm04} given that these lines are observed 
only during the leading edge of the bursts, while the {\em Swift} Burst Alert 
Telescope has an energy passband from 15 to 150 keV.
However {\em Swift} can solve the issue of the hard X--ray lines observed
with {\em Ginga}. A sensitive mission which would be capable to observe the GRB since their
onset and to test the presence of transient X-ray lines with centroid energies below 10 keV
could be the proposed {\em Lobster--ISS} for the International Space Station, an
all-sky monitor with focusing optics from 0.1 to 3.5 keV 
\cite{Fraser02} paired with a GRB monitor for higher photon energies, from 2 keV 
to several hundreds of keV \cite{Frontera03b,Amati04}.

\acknowledgements

We wish to thank Nicola Masetti and Eliana Palazzi for useful
comments.  This research is supported by the Italian Space Agency
(ASI) and Ministry of Instruction, University and Research (COFIN
funds 2001). \sax\ was a joint program of ASI and the Netherlands
Agency for Aerospace Programs.

\clearpage

%
%
\begin{deluxetable}{cccc}
\tablewidth{0pt}
\tablenum{1}
\tablecaption{Spectral parameters of the best fit models}
\tablehead{
Parameters   &     A               &      B       &           C           
} 
\startdata
\multicolumn{4}{l}{{\sc bknpl}}\\
$\Gamma_1$ & $1.5^{+0.2}_{-0.7}$ &  $0.98^{+0.32}_{-0.98}$ & $1.50^{+0.21}_{-0.45}$   \\
$\Gamma_2$  & $1.7^{+0.4}_{-0.2}$ & $1.69^{+0.08}_{-0.07}$ & $2.26^{+2.21}_{-0.24}$   \\
$E_b$ (keV) & $12.4^{+12.0}_{-8.0}$ & $6.9^{+6.1}_{-2.4}$ & $13^{+49}_{-8}$  \\
$\chi^2/{\rm dof}$ & 20.4/22      &  22.1/27           &   16.4/27         \\
\multicolumn{4}{l}{{\sc bknpl} modified by a negative Gaussian (Eq.~\ref{e:model})} \\  
$\Gamma_1$ & $1.2^{+0.4}_{-0.8}$ &  $0.95^{+0.35}_{-1.01}$ & $1.50^{+0.22}_{-0.47}$   \\
$\Gamma_2$  & $1.8^{+0.3}_{-0.1}$ & $1.69^{+0.08}_{-0.07}$ & $2.26^{+2.21}_{-0.24}$   \\
$E_b$ (keV) & $7.4^{+3.5}_{-3.5}$ & $6.9^{+6.5}_{-2.7}$ & $13^{+49}_{-8}$  \\
$E_L$ (keV) & $6.9^{+0.6}_{-0.5}$ &         [6.9]  &    [6.9]               \\
$EW$ (keV)  & $1.2^{+0.5}_{-0.6}$ & $<0.7$ (2$\sigma$) &  $<0.9$ (2$\sigma$)  \\
$\chi^2/{\rm dof}$ & 12.1/20      &  21.7/26           &   16.3/26         \\
\hline
\multicolumn{4}{l}{{\sc bknpl} plus a H--like Fe resonance feature (see text).} \\
$\Gamma_1$   & $1.1^{+0.4}_{-0.8}$ &   --      &	--		\\
$\Gamma_2$   & $1.8^{+0.3}_{-0.1}$ & --	 &	--	       \\
$E_b$ (keV)  & $7.2^{+3.9}_{-3.1}$  &   --      &	--		\\
$N_{\rm Fe\,XXVI + Co\,XXVII}$ (cm$^{-2}$) & $9.4^{+6.9}_{-6.2} \times 10^{19}$ & --    
& --	\\
$\beta \gamma$ & $1.32^{+0.16}_{-0.16}$ & --     &	--	\\
$\chi^2/{\rm dof}$ & 13.1/20              &    --  &      --     \\
\enddata
\label{t:results}
\end{deluxetable}

\clearpage
%
%
\begin{deluxetable}{cc}
\tablewidth{0pt}
\tablenum{2}
\tablecaption{Probability $P_r$, obtained with the run-test, that $n$ consecutive data 
points around the region covered by the depression feature are mutually independent.}
\tablehead{$n$ & $P_r$} 
\startdata
11 &   $2.1\times 10^{-3}$   \\
12  &  $1.2\times 10^{-3}$ \\
13 &   $7.2\times 10^{-4}$  \\
14 &   $2.7\times 10^{-3}$  \\
\enddata
\label{t:rtest}
\end{deluxetable}

%
%
\begin{deluxetable}{ccc}
\tablewidth{0pt}
\tablenum{3}
\tablecaption{Fraction of simulated featureless spectra $f_r$  which provide a 
number of runs with probability $P\le P_r$ (see Table~\ref{t:rtest}) and  fraction 
of spectra $f_t$ which satisfy both the above run test condition and the condition on 
the F-test (see text).}
\tablehead{$n$ & $f_r$ & $f_t$} 
\startdata
11 &  0.0262   &  0.0030  \\
12  &  0.0103 &   0.0009  \\
13 &   0.0044 &   0.0004  \\
14 &   0.0244 &   0.0021  \\
\enddata
\label{t:rftest}
\end{deluxetable}

\clearpage


%
%

%
%

%
%

%
%

%
%

\end{document}